\documentstyle[12pt,tighten,aps,epsfig,amssymb]{revtex}
\preprint{ASI-TPA/10/96}
\title{{\scshape Borel} Quantization: Kinematics and Dynamics}
\author{H.-D.~Doebner and P.~Nattermann}
\address{Institute for Theoretical Physics A,\\
         Technical University Clausthal,\\
         38678 Clausthal--Zellerfeld, Germany,\\ 
         E-mail: {\tt ashdd,aspn@pt.tu-clausthal.de}}
\def\B{{\mathcal B}}
\def\H{{\mathcal H}}
\def\L{{\mathcal L}}
\def\S{{\mathcal S}}
\def\Z{{\mathcal Z}}
\def\rz{{\mathbb R}}
\def\gz{{\mathbb Z}}
\def\cz{{\mathbb C}}
\def\op#1{\mbox{\boldmath $#1$}}        
\def\Ref#1{(\ref{#1})}
\def\X{{\mathfrak X}}
\def\grad{\mbox{grad}}
\def\div{\mbox{div}}
\def\Tr{\mbox{Tr}}
\def\TC{{\cal T}_1^+(\H)}
\def\dt{\frac{\partial}{\partial t}}
\def\splus{\subset \!\!\!\!\!\!\!\, +}
\let\ds\displaystyle
\begin{document}
\maketitle
\pacs{}
\begin{abstract}
  In this contribution we review results on the kinematics of a quantum
  system localized on a connected configuration manifold and
  compatible dynamics for the quantum system including external fields
  and leading to non-linear Schr\"odinger equations for pure states.  
\end{abstract}
\section{Introduction}
Physics starts in general with the notion of space and time. In a 
non-relativistic theory physical objects are understood to be 
localized in space and to move in time. In the case of  
classical mechanics these objects are represented by points in a
configuration manifold $M$ and its building blocks are 
geometrical objects living $M$: {\scshape 
Borel} sets on $M$ or equivalently functions on $M$ may be taken to 
describe the localization of the system, and vectorfields and their
flows on $M$ can be 
used to describe the displacements, i.e.\ the possible movements of 
the system on $M$. In this picture the functions and vectorfields 
serve to describe the kinematics of the classical system; endowed with 
the natural algebraic (semi-direct sum) structure they form the 
\emph{kinematical algebra} $\S(M)$. In symplectic mechanics this algebra 
occurs as the algebra of functions on phase space that are affine in 
momentum. 

The dynamics of such a classical system --- the introduction of time
--- is given by a second order differential equations on $M$, i.e.\
geometrically, by vectorfields on the tangent bundle $TM$ fulfilling a
certain flip condition. If $M$ is (Pseudo-){\scshape Riemann}ian this
condition can be transported to the cotangent  
bundle and yields a natural restriction on the time evolution of
functions on $M$. 

A quantization of the classical theory will therefore involve two 
steps:

First, it requires a representation of the kinematical algebra by 
self-adjoint operators in a separable {\scshape Hilbert}-space. 
Starting from ideas of {\scshape I.E.~Segal} \cite{Segal1} and 
{\scshape G.W.~Mackey}\cite{Mackey1:book} the \emph{Quantum Borel 
Kinematics} was developed \cite{AnDoTo1,Angerm1} and classified
unitarily inequivalent (local and differentiable) representations of 
the kinematical algebra $\S(M)$; we review the results in section  
\ref{QBK:sec}. The phase space description establishes a relation of
this {\scshape Borel} Quantization to Geometric Quantization
(section \ref{rGQ:sec}). This relation will be used in section
\ref{BQef:sec} to generalize the scheme to include external fields.

Secondly, there should be an analogue of the classical condition on 
the time evolution of functions on $M$ for their quantized 
counterparts. This relation will be established in section
\ref{FER:sec}. As we will see in section \ref{NSE:sec} this condition
leads to nonlinear {\scshape Schr\"odinger} equations, if pure states
evolve into pure states. 

\section{Kinematics}\label{Kin:sec}
\subsection{Quantum {\scshape Borel} Kinematics}\label{QBK:sec}
As mentioned in the introduction the idea of Quantum {\scshape Borel}
Kinematics \cite{AnDoTo1,Angerm1} is to describe the quantization of the {\em
localization} and the {\em displacement} of a system on a smooth connected
configuration-manifold $M$.  

The localization is characterized classically by
{\scshape Borel} sets $B\in\B(M)$ and is ``quantized'' by a projection
valued measure   
\begin{equation}
  \label{pvm}
  \op{E}: \B(M) \to \mbox{Proj}(\H)
\end{equation}
on a separable {\scshape Hilbert} space $\H$. Obviously, these
projection valued measures provide a representation of the
infinite dimensional algebra $C^\infty(M)$ of smooth functions on $M$
via the spectral integral
\begin{equation}
  \label{Qf}
  \op{Q}:C^\infty(M)\ni f \mapsto \op{Q}(f) := \int_M f(m) d\op{E}_m
\end{equation}
on a domain $\vartheta_f = \left\{\psi\in\H \left| \int_M |f(m)|^2
    d(\psi,\op{E}_m\psi) < \infty \right\}\right.$.

The classical displacements of the system on $M$ are described by the
flow $\Phi^X_s$ of complete smooth vectorfields
$X\in\X_c(M)$. {\scshape Borel} sets are displaced by 
\begin{equation}
  \label{displace}
  B^\prime_s = \left.\left\{ \Phi^X_s(m) \right| m\in B\right\} \equiv
  \Phi_s^X(B) \in \B(M)\,. 
\end{equation}
\begin{figure}[bt]
\begin{center}
\epsfig{file=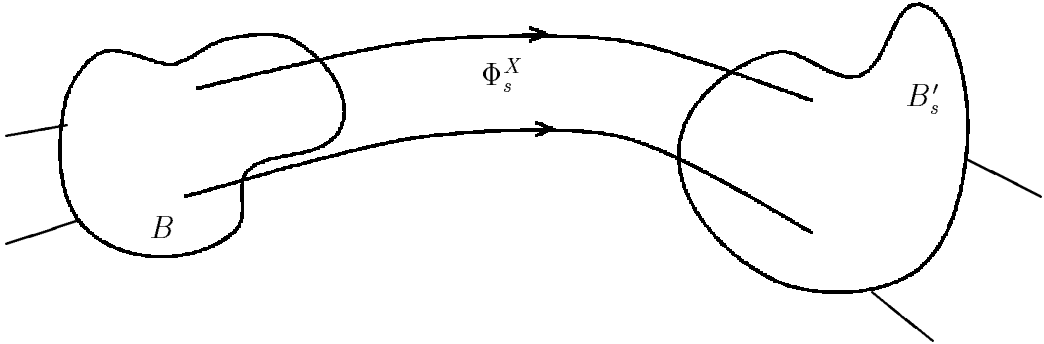}
\end{center}
\caption{Displacement of {\scshape Borel} sets}
\end{figure}

A quantization of these displacements is achieved by a
representation of the flows $\Phi^X_s$ by one parameter groups of
unitary operators on $\H$:
\begin{equation}
  \label{unitary}
  \op{V}_s^X = \exp\left(\frac{i}{\hbar}\op{P}(X)\right)
\end{equation}
with generators $\op{P}(X)$ depending on the vectorfield. The
representation should be consistent with the quantization of the
localization of the system, i.e.~for any given $X\in \X_c(M)$ we require 
$(\op{E},\op{V^X})$ to be a {\em system of imprimitivity}
\cite{Mackey1:book}: 
\begin{equation}
  \label{loc-cons}
  \op{V}_s^X \op{E}(B)\op{V}_{-s}^X = \op{E}\left(\Phi_s^X(B)\right)\,.
\end{equation}

Using fundamental results on the structure of these systems of
imprimitivity \cite{Varada1:book} one can show that for any given
$X\in \X_c(M)$ on a common dense domain $\vartheta_X$ for all $f,g\in
C^\infty(M)$ and $\alpha\in\rz$ \cite{Angerm1}:
\begin{eqnarray}
  \label{Qlin}
  \op{Q}(f)+\alpha \op{Q}(g)&=& \op{Q}(f+\alpha g)\,,\\
  \label{Qcom}
  \left[\op{Q}(f), \op{Q}(g)\right] &=& 0\,, \\
  \label{PQcom}
  \left[\op{P}(X),\op{Q}\right] &=& \frac{\hbar}{i} \op{Q}(\L_Xf)\,. 
\end{eqnarray}
Thus it is natural to assume that the map $\op{P}: \X_c(M)\to \L(\H)$
respects also the algebraic structure of $\X_c(M)$, i.e.\ for all
$X,Y\in\X_c(M)$ and $\alpha\in\rz$ such that $X +\alpha Y,\,[X,Y]
\in\X_c(M)$, respectively, we require
\begin{eqnarray}
\label{partadd}
    \op{P}(X)+\alpha\op{P}(Y)&=& \op{P}(X+\alpha Y)\,, \\ 
\label{parhom}
  \left[\op{P}(X),\op{P}(Y)\right] &=& \frac{\hbar}{i}\op{P}([X,Y])\,,
\end{eqnarray}
together with (\ref{Qlin}--\ref{PQcom}) on a common invariant domain
$\vartheta$. 
Note, that $\X_c(M)$ contains the closed {\scshape Lie}-subalgebra of
vectorfields with compact support $\X_0(M)$ correspond to a {\scshape
  Lie}-algebra homomorphism. 

If we extend the representation of complete vectorfields to {\em all}
vectorfields on $M$ (thereby possibly loosing the selfadjointness of
$\op{P}(X)$), the pair $(\op{Q},\op{P})$ forms a
symmetric irreducible representation of the {\em kinematical algebra}
\begin{equation}
  \label{kA}
  \S(M)=\X(M)\splus C^\infty(M)
\end{equation}
with commutator ($\L_X$ denotes the {\scshape Lie} derivative)
\begin{equation}
\label{kA:comm}
  \Bigl[\left(X_1,f_1\right),\left(X_2,f_2\right)\Bigr]_{ \S(M)} :=  
  \left(\left[X_1,X_2\right]_{\X(M)},\L_{X_1}f_2-\L_{X_2}f_1\right) \,.  
\end{equation}

With further assumptions on the homogeneity of $\op{E}$ (leading
to spinless particles), on the locality of $\op{P}(X)$ and on
$\vartheta$ (leading to
finite differential operators $\op{P}(X)$ w.r.t.\ a differentiable
structure on $M\times\cz$)  
the {\scshape Borel} quantizations $(\op{Q},\op{P})$ have been
classified in \cite{AnDoTo1,Angerm1}:
Unitarily inequivalent representations can be labeled by elements of  
\begin{equation}
  \label{class}
    \fbox{$\ds \pi_1(M)^*\times\rz\,,$}
\end{equation}
where $\pi_1(M)^*$ denotes the group of characters of the fundamental
group of $M$. Furthermore, the {\scshape Hilbert} space can be
realized as the space of square integrable
sections of a flat {\scshape Hermite}an line bundle $(\eta,\langle
.,. \rangle, \nabla)$ with respect to a smooth measure $\mu$ on $M$,     
\begin{equation}
\label{L2sec}
  \H \simeq L^2\left(\eta,\langle.,.\rangle,\mu\right)\,.
\end{equation}
In this realization the representation of the kinematical algebra
reads for all sections $\sigma\in\vartheta\subset
L^2\left(\eta,\langle.,.\rangle,\mu\right)$ ($\chi_B$ denotes the
characteristic function of the set $B$, $\div_\mu$ the divergence
w.r.t.\ the measure $\mu$) 
\begin{eqnarray}
\label{E}
  \op{E}(B)\sigma &=& \chi_B\cdot\sigma \\
\label{Q}  
  \op{Q}(f)\sigma &=& f\cdot \sigma \\
\label{P}
  \op{P}(X)\sigma &=& \frac{\hbar}{i}\nabla_X\sigma +
    \left(c+\frac{\hbar}{2i}\right)\div_\mu X\cdot\sigma\,.  
\end{eqnarray}
In \Ref{class} the elements of $\pi_1(M)^*$ classify the inequivalent
{\scshape Hermite}an line bundles $\eta$ with flat {\scshape Hermite}an
connection $\nabla$ and hence a differentiable structure, whereas
$c\in \rz$ is an additional parameter
independent of the topology of $M$. For a trivial bundle $\eta$ the
{\scshape Hilbert} 
space \Ref{L2sec} is isomorphic to $L^2(M,\mu)$, and \Ref{P} transforms
to ($\psi\in\vartheta\subset L^2(M,\mu)$):
\begin{equation}
  \label{P2}
    \op{P}(X)\psi = \frac{\hbar}{i}\L_X\psi + \omega(X)\psi
    +\left(c+\frac{\hbar}{2i}\right)\div_\mu X\cdot\sigma\,, \\ 
\end{equation}
with a closed real differential one-form $\omega\in Z^1(M)$.

\subsection{Relation to geometric quantization}\label{rGQ:sec}
The quantization method known as \emph{geometric quantization} usually
starts with a general classical 
phase space, i.e.\ a symplectic manifold $(P,\omega)$. Given a
configuration space $M$, the natural choice is 
the cotangent bundle $P:=T^*M$ with canonical one-from
$\theta$ and symplectic form $\omega=d\theta$. A ``full
quantization'', i.e.\ an irreducible selfadjoint representation of the
{\scshape Poisson}-algebra $\left(C^\infty(P),\{.,.\}\right)$ defined
by the symplectic structure fails in general\footnote{For a
  counterexample of a symplectic manifold providing a full
  quantization see \cite{Gotay1}.}. Depending on the polarization
chosen only a sub-algebra of ``quantizable observables'' is
represented irreducibly.  

If, for instance, a complex polarization on $P=\rz^2=T^*\rz$ is
chosen, the sub-algebra $Q(P)$ of polynomials in $x,p$ of max.\ second
order can be represented irreducibly. This leads to a quantization of
the one-dimensional harmonic oscillator including the {\scshape
  Hamilton}ian of the system. Thus the dynamics of the particular
system is also quantized. 

In general for $P=T^*M$ and the vertical polarization the set $L(P)$ of
functions linear in the momenta is used. $L(P)$ is isomorphic to the
kinematical algebra, $L(P)\simeq \S(M)$,
\begin{equation}
  \label{LP-SM}
  \S(M)\ni (f,X) \mapsto Q_f+P_X \in L(P)\,,  
\end{equation}
where the functions $Q_f$ and $P_X$ are defined as 
\begin{equation}
  \label{qf:pX}
  Q_f(\alpha) := f\left(\pi_{T^*M}(\alpha)\right)\,,\qquad
  P_X(\alpha) := \alpha\left(X_{\pi_{T^*M}(\alpha)}\right),\quad\forall
  \alpha\in T^*M\,,
\end{equation}
with {\scshape Poisson} brackets
\begin{equation}
  \label{wcom}
  \left\{Q_f,Q_g\right\} = 0\,,\qquad 
\left\{P_X,Q_g\right\} = Q_{\L_Xf}\,,\qquad 
\left\{P_X,P_Y\right\} = P_{[X,Y]}\,.  
\end{equation}
In geometric quantization only the ``standard'' representation ($c=0$,
$\nabla \hat{=}\L$) of this algebra is considered.

Note that so far only the kinematics of the system has been quantized and
a dynamics has to be introduced by an additional argument. 
In geometric quantization this is generally achieved by choosing
another suitable polarization in which the {\scshape Hamilton}ian is
quantizable and the use of the BKS-kernel (see e.g.~\cite{Sniaty1:book}).
Here we will proceed differently in section \Ref{Dyn:sec}. 

\subsection{{\scshape Borel} Quantization with external
  fields}\label{BQef:sec} 
To describe the interaction with external (magnetic) fields on
$M$ we utilize the phase space picture (see previous section). 
We introduce external fields in terms of closed two-forms $\phi\in
A^2(M)$ on $M$ by changing the symplectic form \cite{Sniaty1:book} on
$T^*M$: 
\begin{equation}
  \label{we}
  \omega_e := d\theta + e\pi^* \phi\,,
\end{equation}
where $\pi:T^*M\to M$ is the projection of the cotangent bundle and
$e$ is a coupling constant (charge).
Using the {\scshape Poisson} bracket $\{.,.\}_e$ induced by this
structure we obtain commutation relations different from \Ref{wcom}:
\begin{equation}
  \label{wecom}
  \left\{Q_f,Q_g\right\}_e = 0\,,\qquad 
\left\{P_X,Q_g\right\}_e = Q_{\L_Xf}\,,\qquad 
\left\{P_X,P_Y\right\}_e = P_{[X,Y]} + e Q_{\phi(X,Y)}\,.  
\end{equation}

A {\scshape Borel} quantization of this algebra leads to the same
operators (\ref{E}--\ref{P}) on
$L^2\left(\eta,\langle.,.\rangle,\mu\right)$, but the {\scshape
  Hermite}an line bundle $\eta$ is not longer flat,  
the external field comes in as a curvature two-form of the bundle
\cite{Drees1}, 
\begin{equation}
  \label{curv}
  R(X,Y) :=  \left[\nabla_X,\nabla_Y\right]-\nabla_{[X,Y]} =
  \frac{ie}{\hbar}\phi(X,Y)\,. 
\end{equation}
However, such line bundles with curvature $R$ exist --- due to a
geometric obstruction --- if and only if  
\begin{equation}
  \label{H2}
  \left[\frac{1}{2\pi i} R\right] \in H^2(M,\gz)\,, 
\end{equation}
i.e.\ the integral of $\phi$ over any singular two-cycle has to be an
integer multiple of $2\pi\hbar$ (see e.g.\ \cite{SimWoo1:book}). 
Hence, only ``quantized'' values of the external field are admissible.  
  
\subsection{Examples}\label{ex1}
To illustrate the method of {\scshape Borel} Quantization we give two
simple examples.
\begin{enumerate}
\item {\scshape Euclid}ean space $M=\rz^n$. \cite{Angerm1,GoMeSh1}\\
  The classification \Ref{class} reduces to $\rz$ since 
  \begin{equation}
    \label{pi-rz}
    \pi_1(\rz^n)=0\,;
  \end{equation}
  in global coordinates $\vec{x}=(x^1,\ldots,x^n)$ the
  vectorfields are (using summation convention) 
  \begin{equation}
    \label{X-Rn}
    X =  X^j(\vec{x})\frac{\partial}{\partial x^j}\,.
  \end{equation}
  The {\scshape Hermite}an line bundle $\eta$ and the connection
  $\nabla$ are trivial and  
  \begin{equation}
    \label{PX-Rn}
    \op{P}(X) = \frac{\hbar}{2i} \left(
      X^j(\vec{x})\frac{\partial}{\partial x^j}+
      \frac{\partial}{\partial x^j} X^j(\vec{x}) \right) + c
    \frac{\partial X^j(\vec{x})}{\partial x^j} 
  \end{equation}
  on $L^2(\rz^n,d^nx)$.
  Note that the extra term does not influence the linear
  and angular momenta 
  \begin{eqnarray}
    \label{Pj}
    \op{P}_j &:=& \op{P}\left(\frac{\partial}{\partial x^j}\right) =
    \frac{\hbar}{i} \frac{\partial}{\partial x^j}\,,\\
    \label{Ljk}
    \op{L}_{jk} &:=& \op{P}\left(x^j\frac{\partial}{\partial x^k} -x^k
      \frac{\partial}{\partial x^j}\right) =
    \frac{\hbar}{i}\left(x^j\frac{\partial}{\partial x^k} -x^k
      \frac{\partial}{\partial x^j}\right)\,.   
  \end{eqnarray}

  Since $\rz^n$ is simply connected, any external field $\phi$ is
  admissible. 
\item The $n$--torus $M=T^n=\underbrace{S^1\times\ldots\times S^1}_{n
    \mathrm{ times}}$ \cite{DoeTol2,Schult1}.\\
  The classification of inequivalent QBK is 
  \begin{equation}
    \label{cl-Tn}
    U(1)^n \times \rz\,,
  \end{equation}
  since 
  \begin{equation}
    \label{pi-Tn}
    \pi_1(T^n) = \underbrace{\gz\oplus\cdots\oplus\gz}_{n \mbox{ times}}\,.
  \end{equation}
  However, the {\scshape Hermite}an line bundles $\eta$ are trivial,
  so that in local coordinates
  $\vec{\varphi}=(\varphi_1,\ldots,\varphi_n)$  
  \begin{eqnarray}
    \label{X-Tn}
    X &=&  X^j(\vec{\varphi})\frac{\partial}{\partial \varphi^j}\,,\\
    \label{PX:Tn}
    \op{P}(X) &=& \frac{\hbar}{i} \frac{1}{2} \left(
      X^j(\vec{\varphi})\frac{\partial}{\partial \varphi^j}+
      \frac{\partial}{\partial \varphi^j} X^j(\vec{\varphi}) \right) + c
    \frac{\partial X^j(\vec{\varphi})}{\partial \varphi^j}+\theta_j
    X^j(\vec{\varphi})\,
  \end{eqnarray}
  on $L^2(T^n,d^n\phi)$ where $\theta_j$ can be chosen to be a
  constant $\theta_j\in[0,2\pi)$. 
  Hence we have a ``topological influence'' on the ``angular'' momentum
  operators:  
  \begin{equation}
    \label{Jj}
    \op{J}_j := \op{P}\left(\frac{\partial}{\partial \varphi^j}\right)
     =\frac{\hbar}{i} \frac{\partial}{\partial \varphi^j}+
      \theta_j\,,
  \end{equation}
  leading to a {\scshape Aharonov-Bohm}-type effect.

  The topology of $T^n$ also restricts the possible choice of the
  external field $\phi$. For constant external fields
  $\phi(X,Y)=\phi_0$, for instance, condition \Ref{curv} implies a ``modified
  {\scshape Dirac} quantization condition'' 
  \begin{equation}
    \label{dirac}
    \phi_0 = \frac{\hbar}{2\pi e} n\,,\qquad n\in\gz\,. 
  \end{equation}
\end{enumerate}

\section{Dynamics}\label{Dyn:sec}
\subsection{Generalized first {\scshape Ehrenfest} relation}\label{FER:sec}
In order to find conditions for the evolution of the quantum system,
let us take a look at the classical system first (see
\cite{Natter1,DoeHen1}). 

Classical dynamics on $M$ (as a configuration manifold) is in
general described by a second order differential equation, i.e.\ a
vectorfield $\tilde{D}$ on the tangent bundle $TM$ fulfilling the
flip condition 
\begin{equation}
  \label{flip}
  T\pi_{TM} \circ \tilde{D} = \mbox{id}_{TM}\,,
\end{equation}
or in local coordinates $(\vec{x},\vec{v})$ of $TM$, 
\begin{eqnarray*}
  \tilde{D}_{(x,v)} \equiv (\dot{\vec{x}},\dot{\vec{v}}) =
  (\vec{v},\vec{F}(\vec{x},\vec{v}))\,\\ 
  \Rightarrow \quad \ddot{\vec{x}} = \vec{F}(\vec{x},\vec{v})\,.
\end{eqnarray*}
If $M$ is (Pseudo-){\scshape Riemann}ian with metric $g$\footnote{For
  an $n$ particle system $g$ could be inherited from the geometry of
  the space manifold and the mass matrix of the 
  particles.} 
there is a
natural isomorphism $g^\flat: TM \to T^*M$ with inverse $g^\sharp$. 
Using this isomorphism we can define 
a dynamical vectorfield $D := Tg^\flat\circ \tilde{D}\circ g^\sharp$
on $T^*M$ and the condition \Ref{flip} turns into
\begin{equation}
  \label{flip-D}
  T\pi_{T^*M}\circ D = g^\sharp\,.
\end{equation}
Let $\Phi_t$ be the flow of $D$ on $T^*M$ and 
\begin{equation}
  \label{cm-evo}
  t\mapsto \alpha_t:=\Phi_t(\alpha_0)
\end{equation}
be the {\em classical time evolution} of the classical state
$\alpha_0\in T^*M$ then from \Ref{qf:pX} and \Ref{flip-D} we get the
following 
condition for the time evolution of the (quantizable) observable $Q_f$
\cite{Angerm1,Natter1,DoeHen1}:
\begin{equation}
  \label{cm-Ehr}
  \fbox{$\ds
  \frac{d}{dt} Q_f(\alpha_t) = P_{\grad_g f}(\alpha_t)\,.
  $}
\end{equation}

We use this formula for a quantum analogue, i.e.\ a condition for the
time evolution of the quantum mechanical states $\Z$,
\begin{equation}
  \label{qm-evo}
  t\mapsto \Z_t := \Phi_t^{QM}(\Z_0)\,,
\end{equation}
and require that in the mean quantum operators behave under the time
evolution of the quantum states $\Z_t$ like the
classical observables under the time evolution of the classical states
$\alpha_t$, i.e.\ a generalization of the
first {\scshape Ehrenfest} relation of quantum mechanics for all $f\in
C^\infty(M)$
\begin{equation}
\label{qm-Ehr}
  \fbox{$\ds 
   \frac{d}{dt}\mbox{Exp}_{\Z_t} \left(\op{Q}(f)\right)  = \mbox{Exp}_{\Z_t}
  \left(\op{P}(\grad_g f)\right)\,. $} 
\end{equation}
Now there are unitarily inequivalent representations of the operator
$\op{P}(X)$ leading to different conditions \Ref{qm-Ehr} on the
time evolution of the quantum mechanical state. 
Indeed, there is {\em no unitary} time evolution of the states
satisfying \Ref{qm-Ehr} except for $c=0$. This is basically due to the
algebraic (multiplicative) structure of $C^\infty(M)$. 

We have two alternative ways to obtain evolution equations satisfying
\Ref{qm-Ehr} for $c\neq 0$. 
 
The first way is to look for evolutions of density matrices, i.e.\ of
positive trace-class operators on $\H$ with trace 1
\begin{equation}
  \label{dm-evo}
  t\mapsto \op{W}_t:= \op{\Phi}^{QM}_t(\op{W}_0)\in\TC \,,
\end{equation}
fulfilling
\begin{equation}
  \label{dm-Ehr}
  \fbox{$\ds
   \frac{d}{dt}\Tr\left(\op{Q}(f) \op{W}_t\right) = \Tr\left(\op{P}(\grad_g
    f)\op{W}_t\right) \,.$}
\end{equation}
By the usual interpretation of $\op{W}\in\TC$ as statistical mixtures,
$\op{\Phi}^{QM}$ has to be {\em linear}.  

For completely positive, norm-continuous\footnote{Actually,
  norm-continuity is a strong restriction. For unitary evolutions it
  corresponds to bounded {\scshape Hamilton}-operators!} there is a
classification of their generators given by {\scshape Lindblad}
\cite{Lindbl1}. Though we are not necessarily looking for
norm-continuous evolutions, there are  
{\scshape Lindblad}-type evolution equations satisfying \Ref{dm-Ehr}
\cite{Natter1,DoeHen1}; 
the details of this way will be explained elsewhere.

The second alternative is to restrict \Ref{qm-Ehr} to pure states,
\begin{equation}
  \label{ps-evo}
  t\mapsto \sigma_t\in L^2\left(\eta,\langle.,.\rangle,\mu\right)\,,
\end{equation}
fulfilling
\begin{equation}
  \label{ps-Ehr}
  \fbox{$\ds
   \frac{d}{dt}\langle \sigma_t\vert\op{Q}(f)\sigma_t\rangle = \langle
  \sigma_t\vert\op{P}(\grad_g f)\sigma_t\rangle  \,.$}
\end{equation}
If the line bundle is trivializable, this leads formally to nonlinear
{\scshape Schr\"odinger} equations for wavefunctions $\psi_t\in
L^2(M,\mu)$, as we will see in the next section. 

\subsection{Nonlinear {\scshape Schr\"odinger} equations}\label{NSE:sec}
For trivial line bundles $\eta$,
$L^2\left(\eta,\langle.,.\rangle,\mu\right)\simeq 
L^2(M,\mu)$, and for wave functions $\psi_t\in L^2(M,\mu)$ condition
\Ref{qm-Ehr} reads   
\begin{eqnarray}
  \frac{d}{dt}\langle \psi_t|\op{Q}(f)\psi_t\rangle &=&\langle
  \psi_t\op{P}(\grad_g f) \psi_t\rangle\\
\label{pure-Ehr}
  \Leftrightarrow\quad \frac{d}{dt}\int_M f(m) \rho_t(m) d\nu_g(m) 
  &=&   \int_M f(m)\left(-j_t(m)+c\Delta_g\rho_t(m)\right) d\nu_g(m)\,,
\end{eqnarray}
where 
\begin{equation}
  \rho_t(m):=\psi_t(m)\bar\psi_t(m),\quad
  j^\omega_t(m):=\frac{\hbar}{i}\Big(\bar\psi_t(m)(\grad_g\psi_t)(m)
  -\psi_t(m)(\grad_g\bar\psi_t)(m)\Big) +\rho_t(m)g^\sharp \omega
\end{equation}
are the probability distribution and the probability current,
respectively. 
Since \Ref{pure-Ehr} has to hold for all $f\in C^\infty(M)$, we get a
{\scshape Fokker-Planck}-type equation 
\begin{equation}
  \label{fpe}
  \dot{\rho}_t +\div^\omega_g j_t = c\Delta_g\rho_t\,.
\end{equation}
restricting the evolution equation of the pure state $\psi_t$
({\scshape Schr\"odinger} equation) to 
\begin{equation}
  \label{nse}
  i\hbar\dt \psi_t = \left(-\frac{\hbar^2}{2}\Delta^\omega_g +V\right) \psi_t
  + i\frac{\hbar c}{2} \frac{\Delta_g\rho_t}{\rho_t}\psi_t 
  + R[\psi_t]\psi_t\,,
\end{equation}
where $\Delta_g^\omega:=\left(\div_g +\frac{i}{2\hbar} \omega\right)
\circ \left(\grad_g +\frac{i}{2\hbar}g^\sharp\omega\right)$, $R[.]$
is some real-valued functional on $\H$ and $V$ is a real 
potential on $M$. 

If we assume for a fairly general  model \cite{DoeGol1,DoeGol4} that
$R[.]$ is ``similar'' to the imaginary nonlinearity
$\frac{\Delta_g\psi}{\psi}$ we get 
\begin{equation}
  \label{Rj}
  \begin{array}{c}
  \ds R[\psi] = \sum_{j=1}^5 R_j[\psi]\,,\qquad\mbox{where}\\[1mm]
  \ds  R_1[\psi] := \frac{\div_g j^\omega}{\rho}\,,\qquad
  R_2[\psi] := \frac{\Delta_g\rho}{\rho}\,,\qquad
  R_3[\psi] := \frac{g(j^\omega,j^\omega)}{\rho^2}\,, \\[1mm]
\ds  R_4[\psi] := \frac{d\rho\cdot j^\omega}{\rho^2}\,,\qquad 
  R_5[\psi] := \frac{d\rho\cdot\grad_g\rho}{\rho^2}\,.
  \end{array}
\end{equation}

\section{Final remarks}
We have shown how nonlinear quantum mechanical
evolution equations arise from {\scshape Borel} quantization on a connected,
(Pseudo-){\scshape Riemann}ian configuration manifold; they are
natural generalizations of the {\scshape Doebner-Goldin} equations on
$M=\rz^3$ \cite{DoeGol1,DoeGol4} to more general manifolds. 
Some of the properties (see e.g.~the contributions in
\cite{DoDoNa1:proc}) of the {\scshape Doebner-Goldin} equations in $\rz^3$ are
also valid on more general {\scshape Riemann}ian manifolds $M$. 

In particular, there is a sub-family linearizable \cite{Natter2}
by\footnote{Note that for $\Lambda\neq \pm 1$ the linearization is
  well-defined only for non-vanishing wave-functions.}
\begin{equation}
  \label{gt}
  \psi \mapsto N_{(\Lambda,\gamma)}[\psi] :=
  \psi^{\frac{1}{2}(1+\Lambda+i\gamma)}
  \bar\psi^{\frac{1}{2}(1-\Lambda+i\gamma)} \,. 
\end{equation}
Obviously, this transformation leaves the probability density
invariant. As in
non-relativistic quantum mechanics all measurements can in principle be
reduced to positional ones (see e.g.~\cite{FeyHib1:book,Mielni1}), 
$N_{(\Lambda,\gamma)}$ was thus called a
\emph{nonlinear gauge transformation} and one can identify gauge
equivalent classes among the equations in \Ref{nse} \cite{DoGoNa1}.

Because of some confusion in the context of nonlinear quantum theories
(and superluminal communications therein) we emphasize finally that
equation \Ref{nse} describes only the nonlinear time evolution of pure
states. Mixtures of these pure states have to be identified according
to the set of observables. A description of this 
idea has already been given by \textsc{B.~Mielnik} in \cite{Mielni1}:
taking positions as \emph{primitive observables} and generating the
set of all observables by combining the primitive observables with the
time evolutions (under different external conditions such as $V$ and
$\phi$) one defines a mixture as an equivalence class of probability
measures on the set of pure states w.r.t.\ the observables. By
construction, the so-defined mixtures are consistent with the time
evolution of pure states and it is evident for a nonlinear
time evolution of pure states that the mixtures are \emph{not}
represented by trace class operators in $\TC$ (see also \cite{Luecke1}
for a discussion of observables in a nonlinear theory).
\subsection*{Acknowledgments}
The authors are grateful to the German-Polish Foundation whose support
made this Polish-German and international symposium
possible. Especially, we acknowledge discussions with our Polish
colleagues, as well as {\scshape G.A. Goldin}, {\scshape W.~L\"ucke},
and {J.D.~Hennig}.

\end{document}